\documentstyle[amstex,psfig]{article}

\newcommand{\real}{{\Bbb R}}

\newcommand{\natnum}{{\Bbb N}}

\newtheorem{theorem}{Theorem}
\newtheorem{lemma}[theorem]{Lemma}

\newcommand{\cinf}{C^{\infty}}
\newcommand{\dgleich}{\stackrel{\mbox{\tiny {def}}}{=}}
\newcommand{\comm}[2]{[#1 \stackrel{\raisebox{-2pt}[0mm][0mm]{$\wedge$}}{,}
                       #2]}

\newcommand{\tensor}{\otimes}
\newcommand{\EXP}{\mbox{EXP}}
\newcommand{\fib}{\mbox{\scriptsize fib}}
\newcommand{\jEXP}{j^\infty_p(\EXP_p)}
\newcommand{\Gl}{\mbox{Gl}}
\newcommand{\restr}[1]{\;\rule[-2mm]{0.2mm}{5mm}_{\;#1}}
\newcommand{\tot}{\mbox{\scriptsize tot}}
\newcommand{\comment}[1]{}

\newcommand{\cala}{{\cal A}}
\newcommand{\starh}{*_{\hbar}}

\title{{\bf The differential geometry of Fedosov's
quantization}}
\author{Claudio Emmrich\thanks{Research supported by Deutsche
Forschungsgemeinschaft,
Az.: Em 47/1-1.}\mbox{~}  and
Alan Weinstein\thanks{Research partially supported by NSF
Grants DMS-90-01089 and DMS-93-09653.}
\\Department of Mathematics\\
University of California\\
Berkeley, CA 94720 USA\\
{\tt\small(emmrich@@math.berkeley.edu,alanw@@math.berkeley.edu)}}
\begin{document}
\maketitle
\begin{abstract}
B. Fedosov has given a
simple and very natural construction of a deformation quantization for
any symplectic manifold, using a flat
connection on the bundle of formal Weyl algebras associated to the tangent
bundle of a symplectic manifold.  The connection is obtained by
affinizing, nonlinearizing, and iteratively flattening a given torsion free
symplectic connection.  In this paper, a classical analog
of Fedosov's operations on connections is analyzed and shown to
produce the usual exponential mapping of a linear connection on an
ordinary manifold.  A symplectic version is also analyzed.  Finally,
some remarks are made on the implications for deformation quantization
of Fedosov's index theorem on general symplectic manifolds.
\end{abstract}

\section{Introduction}
\label{sec-intro}
In a remarkable paper \cite{fe:simple}, B. Fedosov
presented a
simple and very natural construction\footnote{Actually, this
construction appeared in several earlier papers, such as
\cite{fe:formal} and \cite{fe:index}.} of a deformation quantization for
any symplectic manifold.  The construction begins with a linear
symplectic connection on the tangent bundle of a manifold and proceeds
by iteration to produce a flat connection on the associated bundle of
formal Weyl algebras.  The aim of this paper is to give some geometric
insight into Fedosov's construction by exploring some of its purely classical
analogs.  Our basic idea is that Fedosov's quantization
procedure involves a sort of ``quantum exponential mapping.''

The paper is divided into two parts.  The first part
(Sections \ref{sec-deformation} through \ref{sec-further})
is mostly discursive and follows fairly closely the lecture given by
the second author at the conference in honor of Bert Kostant's 60'th
birthday.  The second part (Sections  \ref{sec-construction} through
\ref{sec:sympl}), more technical and
logically self-contained, develops the proofs of some of the
statements in the first part.

\section{Deformation quantization and symplectic connections }
\label{sec-deformation}

The basic deformation quantization problem is to put a noncommutative
(associative) product structure on the formal power series ring $\cala
[[\hbar]]$ over a commutative Poisson algebra $\cala $.  The physical
origins of this particular formulation of the classical limit are
obscure to this author (see Section 16.23 in \cite{bo:quantum}
for an occurrence in a textbook), but the
origins of its recent intensive study seem to lie in the work of Berezin
\cite{be:quantization} and Bayen {\em et al} \cite{bffls:deformation}.

The deformed product, which is determined by the product of elements
of $\cala $ and can be written in the form $$ f\starh g =
\sum _{j} \hbar ^{j}B_{j}(f,g)$$ for bilinear operators $B_{j}$ on
$\cala $, is assumed to start with the original commutative product
followed by the Poisson bracket, i.e.,

$$ f\starh g = fg + (i\hbar/2)\{f,g\}+\cdots $$

It was already noted in \cite{bffls:deformation} that, when $\cala $
is the algebra of $C^{\infty}$ functions on a symplectic manifold $P$,
the term of order $\hbar^{2}$ in the deformed product is closely
related to a torsion-free {\em symplectic connection} on $P$.  If
there exists such a connection which is {\em flat}, one can
immediately write down a solution of the deformation quantization
problem.  In fact, $P$ is then covered by coordinate charts for which
the transition maps are affine symplectic transformations, which leave
invariant the Moyal product \cite{bffls:deformation} on $\cinf (\real
^{2n})[[\hbar]]$.  (The algebra with this product is often called the
(smooth) {\bf \em Weyl Algebra}.)

When $P$ does not admit a flat symplectic connection (or where such a
connection cannot be invariant under an important symmetry group, such
as in the case of the hyperbolic plane with its $PSL(2,\real )$
symmetry), it is necessary to ``piece together'' quantizations like
the one above in a more elaborate way.  The first such construction to
work for all symplectic manifolds was that of DeWilde and Lecomte
\cite{de-le:existence} who actually piece together quantizations on
individual coordinate systems with nonlinear transition maps.

A more geometric formulation of the ``piecing construction'' was given
by Maeda, Omori, and Yoshioka in \cite{om-ma-yo:weyl}.  Their
construction has important elements in common with that of
Fedosov, so we will describe it in some detail here.

For any symplectic manifold $P$, we consider its tangent bundle $TP$
with the Poisson structure for which each fibre is a symplectic leaf,
carrying the translation invariant symplectic 2-form associated
with its structure as a symplectic vector space.  It is very simple to
quantize this Poisson manifold --- we simply quantize each fibre by the
Moyal product, so that the noncommutative algebra $\cinf (TM)[[\hbar]]$
may be identified with the set of ``smooth'' sections of a bundle
whose fibre over each point $p$ of $P$ is the smooth Weyl algebra of the
tangent space.  Geometrically, we think of $\cinf (TM)[[\hbar]]$ as the
``space of functions on the quantized tangent bundle''.

To quantize $P$ itself, we may try to embed $\cinf (P)[[\hbar]]$ as a
subalgebra of this noncommutative $\cinf (TP)[[\hbar]]$.  The embedding
onto the functions which are constant on fibres of $TP$ does not work,
since this subalgebra is commutative, so we need a different
embedding.  In geometric terms, we are trying to replace the usual
projection of $TP$ to $P$ by a different one.

To see what this new projection should look like, we may examine the
simplest case, where $P$ is a symplectic vector space.  Denoting by
$x$ some linear coordinates on $P$ and $(x,y)$ the corresponding
coordinates on $TP$, the multiplication on the quantized $TP$ is just
the ``Moyal product in $y,$ with $x$ as a parameter.''
If we identify each function $f(x)$ on $P$ with
the function $Lf(x,y)=f(x+y)$ on $TP$, then taking the fibrewise Moyal
product of $Lf$ and $Lg$ reproduces the ``Moyal product in $x$'' of
$f$ and $g$.  The operator $L$ is just pullback by the map
$(x,y)\mapsto (x+y),$ which is nothing but the {\em exponential
mapping} of $P$ with its flat affine structure.  (Here, the
deformation parameter $\hbar$ just comes along for the ride.)
It is not hard to see that the same idea works for any symplectic
manifold with a flat, torsion--free symplectic connection: pulling
back functions from $P$ to $TP$ by the exponential map allows one to
pull back the multiplication on the quantized tangent bundle to get
the aforementioned Moyal-type deformation quantization of $P$.

At this point, one might object that the connection on $P$ might not
be complete, in which case its exponential map is not globally
defined.  In this case, it suffices to restrict the fibrewise Moyal
product to the open subset of $TP$ which is the domain of the
exponential map.

When the connection on $P$ has nonvanishing torsion or curvature, the
situation is much worse, in that the set of functions pulled back
 from $P$ by the exponential map is not closed under the
multiplication on the quantized $TP$.  Two remedies for this problem have been
used in the literature.  We may describe them roughly as follows.
In \cite{om-ma-yo:weyl}, the local exponential mappings coming from a covering
of $P$ by Darboux coordinate systems are ``patched together''
 to produce a new projection which is no longer (at least not manifestly) the
exponential map of
a connection.  This new projection is built at the ``quantum level'': that is,
what is actually constructed is a patching together of modifications of
the local
 pullback mappings from functions on the base to subalgebras of local
sections of the bundle of Weyl algebras.
In fact, this description is slightly oversimplified.  It is not the bundle of
smooth Weyl algebras but rather the bundle of formal Weyl algebras (formal
power series around zero of functions on the tangent spaces) which must be
used.
Also, the parameter $\hbar$ now plays an essential role, in that the pullback
of a function on the base is no longer independent of $\hbar$.

Fedosov \cite{fe:simple}, on the other hand, beginning with
a linear symplectic connection having zero torsion
but nonvanishing curvature, lifts it to a connection on the bundle of formal
Weyl algebras and then
modifies the connection on this bundle, with structure Lie algebra
 the inner derivations, to make it flat. He then shows how
 the space of parallel sections of this bundle, which is clearly a subalgebra
of the algebra of all sections, may be identified with the space
$\cinf (P)[[\hbar]]$.

The idea which we propose in this paper is that Fedosov's construction can also
be interpreted on the level of spaces.  The parallel sections of the bundle of
formal Weyl algebras are defined by the vanishing of covariant differentiation
operators, which are derivations on the space of sections.  When the space of
sections is considered as the space of functions on the quantized $TP$ (more
precisely, on a formal neighborhood of the zero section in the quantized $TP$),
these
derivations may be thought of as vector fields, determining a distribution
on $TP$ which is transverse to the fibres, i.e. an {\bf \em Ehresmann
connection},
on $TP.$  The flatness of Fedosov's connection is equivalent to the
involutivity of the distribution, or the flatness of the Ehresmann connection.
Thus, the parallel sections of the formal Weyl algebra bundle are interpreted
as functions on the quantized $TP$ which are constant along the leaves of the
quantum foliation which is tangent to the quantum Ehresmann connection.

To pull back a function on $P,$ we first
identify it with a function on the zero section, then extend it to a function
on $TP$ constant along leaves.  This is possible if the Ehresmann
connection is not tangent to
the zero section of $TP,$ but is rather transverse to the zero section, so that
each leaf intersects
the zero section in a unique point.  To achieve this, even if the initial
linear connection is flat, we must modify it by the addition of a
``translation'' term to obtain a connection with affine structure group.

The somewhat mysterious fact that a parallel section
 for the Fedosov connection is not determined by its value at a single point
can also be attributed to the addition of the translational term, which changes
the way in which the covariant differential operator behaves with respect to
the grading in the formal Weyl algebra.  In other terms, the fact is
related to the non-determination of a $\cinf $ function by its Taylor
series at one point.  An analogous situation occurs on the bundle of
infinite jets of real-valued functions on $P$, which carries a ``flat
connection''  for which the parallel sections are the infinite jets of
functions on $P$.  In a sense, what Fedosov's construction does is to
identify this bundle of infinite jets of functions on $P$ with the
bundle of formal functions on the fibres of $TP$ and then to pull back
the natural flat connection from the former bundle to the latter.

There is another way to see the necessity of ``affinizing'' the structure
group of the connection.  If a nonlinear flat connection on $TM$ has the
zero section as a parallel section, then its linearization at the origin
would be a flat linear connection.  There are, in general, obstructions to the
existence of such a connection; by allowing the zero vectors to move, we
bypass these obstructions.

\section{Classical analogs}
\label{sec-classical}

In the previous section, we used geometric language to give an intuitive
picture of the constructions on quantized formal power series in
\cite{fe:simple}.  Now
we will turn the tables and apply Fedosov's formal constructions in purely
classical settings.  A more detailed version of the following remarks is
contained in the second part of this paper, beginning with Section
\ref{sec-construction}.

Suppose that we are simply given a differentiable manifold $P$ with a torsion
free linear connection on $TP$.  This connection may be lifted to a connection
on the associated bundle of algebras of formal power series on the tangent
spaces of $P.$  If we extend the structure group of this connection from
the Lie algebra of linear vector fields on $\real^{m}$ ($m$ being the
dimension of $P$) to the Lie algebra of formal vector fields, then Fedosov's
iteration method can be used to produce, in a canonical way, a flat connection
on this bundle for which the corresponding Ehresmann connection is transverse
to the zero section.  The leaves of the foliation given by the flat Ehresmann
connection are the fibres of a ``mapping'' from a formal neighborhood of the
zero section in $TP$ to $P$.   We denote this formal mapping by $\EXP.$

On the other hand, there is another well known mapping from $TP$ to $P$
determined by a connection, namely the usual exponential mapping $\exp.$  What
is the relation between these two mappings?  Since both are produced in a
canonical manner from the connection, it is natural to guess that they are
equal.  In fact this is true, as we prove below.  The structure of the proof
is of some interest.  We first prove that, for a real-analytic connection,
the flat connection given by Fedosov's iteration is given by convergent
power series, so it actually defines a flat real-analytic Ehresmann connection
on a neighborhood of the zero section.  In this case, we prove by following
geodesics that $\EXP$ and $\exp$ coincide.  Next, we show that $\EXP$ can be
defined in the purely formal context by ``extension by continuity,'' using the
fact that the convergent power series (already the polynomials) are dense in
all the power series with respect to the uniformity for which two power series
are close if all their coefficients up to some high degree are equal.
Finally, we conclude that $\EXP$ and $\exp$ are equal since they agree on a
dense subset.  Since all the constructions are local, they then apply on
any manifold, without any reference to a real-analytic structure.

In a sense, we have shown that the iterative procedure used by Fedosov to
``flatten''
a connection, when applied to ordinary manifold, is simply another way of
constructing the usual exponential mapping.  It is in this sense that we are
tempted to say that
his quantization procedure involves the pullback of functions from a
symplectic manifold $P$ to its quantized $TP$ by quantization of the
exponential mapping of a given symplectic connection.  Before taking this leap,
though,
we must be more careful, because the exponential mapping of a symplectic linear
connection on a symplectic manifold does NOT in general define symplectic
mappings from the tangent spaces with their constant symplectic structures to
the manifold with its given symplectic structure.

To produce from a symplectic torsion-free
linear connection on $TP$ a symplectic nonlinear connection (on a formal
neighborhood of the zero section), we must change the structure Lie algebra
from the algebra of formal vector fields to the algebra of formal symplectic
vector fields.  In fact, with a view toward quantization, it is useful to use
instead the one-dimensional extension of the latter algebra given by the
infinite jets of functions, with the Poisson bracket Lie algebra structure.
As is well known, this Lie algebra acts on itself by derivations of both the
multiplication and the Poisson bracket.  Now the iterative procedure can be
applied again in this symplectic classical context to produce (formal)
symplectic ``exponential'' mappings from the fibres of $TP$ to $P$.
It would be interesting to have a more geometric construction of these
mappings.  Perhaps they are the same as those obtained by starting with the
ordinary exponential mappings and then correcting these by using the
deformation-method proof of Darboux's theorem in \cite{we:symplectic}.

\section{Rigidity of quantization and Fedosov's index theorem}
\label{sec-further}

In the sections above, we have concentrated on Fedosov's construction of
a deformation quantization.  This construction was, however, merely the
beginning of Fedosov's work, one of whose points of culmination is an
index theorem for arbitrary symplectic manifolds
\cite{fe:index}.  This theorem is a
generalization of the Atiyah-Singer theorem, to which it reduces when the
symplectic manifold is a cotangent bundle.

In this section, we wish to comment on the role which Fedosov's index theorem
might play in resolving a fundamental question in the general theory of
deformation quantization.  The condition
$$ f\starh g = fg + (i\hbar/2)\{f,g\}+\cdots $$
which describes the $\hbar$ dependence of the product $\starh$ says nothing
about how the product behaves after the deformation parameter ``leaves the
first-order neighborhood of zero.''  This weak condition seems to contrast
with the rigidity implied by the special role of rational values of
$q=e^{i\hbar}$ in the theory of quantum groups.  The question thus arises
as to whether the dependence on $\hbar$ should be ``rigidified'' by
some supplementary condition(s).

A possible source of such rigidity is suggested in \cite{we:classical}, where
in the special case of quantizations of Moyal type on tori with
translation-invariant Poisson structures it is
shown that the Schwartz kernel of the bilinear multiplication operator
$\starh$ satisfies a Schr\"odinger equation in which
Planck's constant plays the role of time,
and the hamiltonian operator is the Poisson structure
extended by translation invariance to become
a differential operator on $P\times P.$  Although this kind of evolution
equation for quantization may possibly extend to other situations where the
notion of translation makes sense (e.g. on Lie groups), it is not at all
clear how to extend it to general symplectic manifolds.

Another kind of rigidity in the deformation parameter appears in the
geometric quantization of K\"ahler manifolds by sections of
holomorphic sections of line bundles (and the extension of this
quantization to general symplectic manifolds by the method of Toeplitz
operators in \cite{bo-gu:spectral}).  The deformation parameter
$\hbar$ then occurs as a unit by which the cohomology class of the
symplectic structure (which in physical examples carries the units of
action) is divided to obtain a cohomology class with numerical values
which, when integral, is the first Chern class of a complex line
bundle.  When this class is integral for $\hbar=\hbar_0,$ then as
$\hbar$ runs over the set of values $\hbar_0 /k$, for $k$ a
sufficiently large integer, the dimension of the space of holomorphic
sections is given according to the Riemann-Roch theorem by a
polynomial in $k$ of degree equal to half the dimension of the
symplectic manifold, which is completely determined by the cohomology
class of the symplectic structure and the total Chern class of the
tangent bundle for an almost complex structure compatible with the
symplectic structure.  This behavior shows that geometric quantization
has a very rigid behavior with respect to $\hbar.$

The index theory of Fedosov
\cite{fe:index} provides an analog in deformation quantization
of the rigidity described in the preceding paragraph.  According to this
theory, when his deformation quantization is extended from scalar functions
to matrix-valued functions on $P,$ a notion of ``abstract elliptic
operator'' can be defined.  When a suitable trace is introduced, the index of
such an operator can be defined.  Fedosov's index theorem then expresses
this index as a polynomial in $(1/\hbar)$ which is completely determined by
the Chern character of
a certain vector bundle over $P$ associated to the operator, the
$\hat{A}$ class of the tangent bundle of $P$ (with an almost complex
structure), and the cohomology class of the symplectic structure.  Once
again, this formula suggests that some quantizations (for which this
formula is true) are ``better'' than others which might have the same
first-order behavior with respect to $\hbar.$

\section{Construction of formal flat connections}
\label{sec-construction}
In this section we study in detail the classical analog of the flat
connection $D$ constructed in \cite{fe:simple} for the quantum
case. In this classical setting  the Weyl algebra bundle is replaced by
 ${\cal A}={\cal F}(TP)$, the algebra of smooth functions on $TP$, or,
 if one restricts oneself to the study of formal power series on
 the fibres, by ${\cal A} = \Gamma( \bigcup_{p \in P} {\cal J}^\infty_0(
 T_pP,\real))$, where  $ {\cal J}^\infty_0(T_pP,\real)$ denotes the set of
 $\infty$-jets at 0 of real valued functions on $T_p P$. The product
on ${\cal A}$  in the non-symplectic setting is
 either pointwise multiplication  for ${\cal A}={\cal F}(TP)$, or, for
${\cal A} = \Gamma( \bigcup_{p \in P} {\cal J}^\infty_0( T_p P,\real))$,
 the mapping induced by it on the set of $\infty$-jets, respectively.
In the symplectic case studied in Section \ref{sec:sympl} this product
is supplemented by the fibrewise
 Poisson bracket (i.e., the Poisson bracket on the fibres of TP induced by
 the symplectic form $\omega$ on $P$), to produce a Poisson algebra structure.

 Let $\Lambda(P)$ be the set of forms on $P$.  The contraction
$(i_X \alpha)(\cdot)$ is defined as $\alpha(X, \cdot)$ for a vector $X$
and a form $\alpha$.
 As in \cite{fe:simple},  we can define operators $\delta$ and $\delta^{-1}$
 on ${\cal A} \tensor \Lambda(P)$, given in local coordinates $(x^i)$ on $P$
 and induced coordinates $(x^i,y^i)$ on $TP$ by:
 \begin{equation}
  \delta = d x^i \wedge \frac{\partial}{\partial y^i}
 \end{equation}
 and
 \begin{equation} \label{eq:deltainv}
    \delta^{-1} a = \frac{1}{p+q} \; i_{ y^i \frac{\partial}{\partial x^i}} a
 \end{equation}
 if $a$ is a $q$-form that is homogenous of degree $p$ in $y$, $p+q \neq 0$,
 and $\delta^{-1} a =0 $ if $p=q=0$.

 Let $ {\cal V}^{inv}(TP)$ denote the set of invariant vertical vector fields,
 i.e., of vertical lifts of vector fields on $P$,
We can define the operators $\delta$ and $\delta^{-1}$ on forms with values
 in the  `formal vertical vector fields'
 ${\cal A} \tensor  {\cal V}^{inv}(TP) \tensor \Lambda(P)$ as well,
 by letting them
 act trivially on $ {\cal V}^{inv}(TP)$.
 With this definition, the ``Hodge decomposition''
 \begin{equation} \label{eq:delta}
  a =a_{00} + (\delta^{-1} \delta  + \delta \delta^{-1} ) a
 \end{equation}
 still holds for $a \in  {\cal A} \tensor {\cal V}^{inv}(TP)\tensor\Lambda(P)$.
 Here, $a_{00}$ denotes the homogeneous part of $a$ which is a zero-form of
degree 0 in the vertical coordinates $(y^i)$; i.e. it is simply an
element of ${\cal V}^{inv}(TP)$.

 We would like to construct from a given linear, torsion-free, nonflat
 connection $\partial$ a flat, nonlinear connection on $TP$ with structure
 group $\mbox{Diff}(\real^n)$ (or the symplectomorphism group of $\real^n$
for even $n$ in the case of the
Poisson algebra studied in Section \ref{sec:sympl}),
 or equivalently, a flat Ehresmann connection on $TP$.
 Locally, such a connection is represented by a
 one-form $X_{\tot}$ on $P$ with values in the vertical vector fields,
  yielding for the local expression for the
 curvature $\Omega$, i.e., the obstruction to the integrability
 of the horizontal distribution defining the connection:
 \[ \Omega = d X_{\tot} + 1/2 \comm{X_{\tot}}{X_{\tot}} \]
 where $\comm{\cdot }{\cdot }$ acts as the Lie bracket on the vector part and
as the  exterior product on the form part of $X$.
A connection induces a covariant derivative $D$ on elements of ${\cal
A}$, which is given in local coordinates by:
\[ D f =\left( \frac{\partial}{\partial x^i} f\right)  d x^i
     + X_{\tot} \cdot f. \]

By analogy with the quantum case in \cite{fe:simple}, $X_{\tot}$ should
be locally  of the form
 \begin{equation} \label{eq:Xansatz}
  X_{\tot} = -    \frac{\partial}{\partial y^i} d x^i
    - \Gamma^i_{kl}(x) y^k \frac{\partial}{\partial y^i} d x^l    + X ,
 \end{equation}
 where $\Gamma^i_{kl}(x)$ are the Christoffel symbols of the
connection $\partial ,$
 and $X$ is at least of second degree in $y$.

The addition of the term
 $     \frac{\partial}{\partial y^i} d x^i$, which corresponds to the operator
 $\delta$, corresponds to the transition from the linear connection
$\partial $ to an affine connection with connection form $X_{\tot }-X$
by the addition of the solder form \cite{ko-no:foundations}.  The vanishing of
the torsion for $\partial $ implies that the ``translational'' part of
the curvature of the affine connection is zero.

For instance, if the connection $\partial$
is flat, we may choose $X \equiv 0$ in $(\ref{eq:Xansatz})$ to get a flat
connection again. In this case, the original horizontal distribution
corresponding to $\partial$ is already integrable.
For $P= \real^n$ with the canonical flat connection, the leaves of the
horizontal distribution correponding to $\partial$ are affine subspaces
$\real^n \times \{v\} \subset \real^{2 n}$  parallel
 to the zero section
$Z = \real^n \times \{0\} \cong P$. This foliation is rotated by the
addition of the solder form in such a way that the leaf through $ (x,v)$
intersects the zero section in $(x+v,0)$ for arbitrary $(x,v) \in \real^{2n}$
(Figure 1).

\begin{figure}
\centering
\mbox{\psfig{file=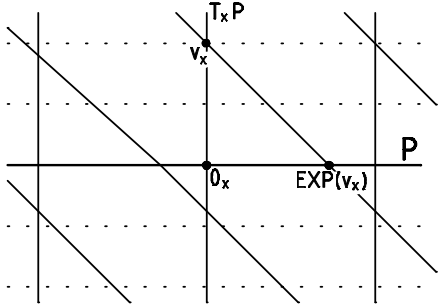,width=6cm,angle=270} }
\caption{Addition of the solder form to a fl at linear connection
``rotates'' the parallel sections (from dotted to solid lines) so that
they become transversal to the zero section.}
\end{figure}

 In the general, non-flat case one may interpret the ansatz
$(\ref{eq:Xansatz})$ as first going over from the structure group
$\Gl(n,\real)$ with $n= \dim(P)$ to the affine group
$\Gl(n,\real) \lhd \real^{n}$ in order to get a horizontal distribution
transversal to the zero section, and then deforming it in order to
get a flat nonlinear connection, which has an integrable horizontal
distribution and hence defines a foliation.

Although the transition from a linear connection to an affine
connection is canonical, one could in principle take any multiple of
$\delta$ in order to define a generalized affine connection in the
sense of \cite{ko-no:foundations}, i.e., a connection on the
$\Gl(n,\real) \lhd \real^{n}$-bundle of affine frames. However, the
choice made in the ansatz is not only geometrically natural, but also
distinguished by the validity of Theorem \ref{thm1} and Lemma
\ref{lem1} below.

 The requirement of the vanishing of the curvature yields the condition:
 \begin{equation} \label{eq:curvzero}
  0 = \Omega = R - \delta X + \partial X +  1/2\comm{X}{X} ,
 \end{equation}
 where  $ \partial X \dgleich  d X -
 \comm{ \Gamma^i_{kl}(x) y^k \frac{\partial}{\partial y^i} d x^l}{X}$,
and $R$ denotes the curvature tensor corresponding to $\partial$.
The term
$\comm{\frac{\partial}{\partial y^j} dx^j}
      {\Gamma^i_{kl}(x) y^k \frac{\partial}{\partial y^i} d x^l}$
which one would expect to appear in (\ref{eq:curvzero}) is equal to
$\Gamma^{i}_{kl}(x) dx^{j}\wedge dx^{l} \frac{\partial}{\partial
y^i},$ which vanishes because the connection is torsion--free.

 We first try to find a formal power series in $y$ for the solution of
 Equation (\ref{eq:curvzero}).  Solutions of this equation are not unique;
however, we will show that they are in one-to-one correspondence to
 formal vertical vector fields
  $a \in {\cal A} \tensor  {\cal V}^{inv}(TP)$ of degree at least
 3 in $y$ via the  `normalization condition'
 \begin{equation} \label{eq:bound}   \delta^{-1} X = a.
 \end{equation}
We will see in Section \ref{sec:anal} that $\delta^{-1} X$ is just the
 $X$-dependent
term that appears in the equation defining autoparallel curves. Hence,
the case $a=0$ will prove to be special in so far as the term involving
 $X$ does not change autoparallel curves: in general $\delta^{-1} X$ can be
interpreted as an acceleration term for autoparallel curves.

Therefore it might appear natural to admit only the special case $a=0$.
However, we will see in section \ref{sec:sympl} that the boundary condition
$\delta^{-1} X$ is not appropriate in the symplectic case.  (It must
be replaced by the condition
$\delta^{-1} r =0$ for the corresponding hamiltonian function $r$.)
Hence, we will admit arbitrary values of $a$ in the non-symplectic case
as well.

Using $(\ref{eq:delta})$ and the vanishing of $X_{00}$ (since $X$ is a
one-form), we get:
 \begin{equation} \label{eq:iter}
  X = \delta a + \delta^{-1}(R + \partial X + 1/2 \comm{X}{X}),
 \end{equation}
which must be satisfied by any solution to $(\ref{eq:curvzero})$
and $(\ref{eq:bound})$.
 As in \cite{fe:simple}, this equation may be solved by iteration,
 yielding a unique power series for $X$. In spite of the more general
 normalization condition $(\ref{eq:bound})$, the argument in
 \cite{fe:simple} (also see the ``topological lemma'' in
\cite{do:quantization}) showing that the $X$ found in this way is indeed a
 solution to $(\ref{eq:curvzero})$  is still valid:
  defining $A$ as the right hand side of equation
 $(\ref{eq:curvzero})$ with the recursively constructed $X$, we may
check that $A$ fulfills the
 equation $ \delta A = \partial A +[X,A]$
  and the condition $\delta^{-1} A= 0$ and hence vanishes
 identically.

Hence, we have proved the following theorem
\begin{theorem}
For each  $a \in {\cal A} \tensor  {\cal V}^{inv}(TP)$ at least of degree
 3 in $y$ there is a unique formal flat connection, i.e., a
solution to $(\ref{eq:curvzero})$, satisfying the normalization
condition $(\ref{eq:bound})$.
\end{theorem}

\section{Analytic connections and the ``exponential map''} \label{sec:anal}
In the last section we constructed a formal flat connection from a
given linear, torsion-free connection. In this section we will show
that this formal power series is convergent in a tubular neighborhood $U$
of the zero section if the manifold $M$, the connection $\partial$, and
the element $a$ in the normalization condition are real analytic.  The
proof uses the method of majorants, as discussed in any standard
reference on real analytic functions, such as \cite{kr-pa:primer}.

 \begin{theorem} \label{thm0}
 Let $P$ be a real analytic manifold, $\partial$ an analytic linear
 connection, and $a \in {\cal F}(TP)$ an analytic function
 such that $a$ and its derivatives up to order two
 in the direction of the fibres vanish on
the zero section $Z$. Then the
 formal power series which is the solution of $(\ref{eq:iter})$
obtained by iteration is convergent in a
 neighborhood $U$ of $Z$  and hence defines an analytic
 Ehresmann connection on $ U \subset TP$.
 \end{theorem}

 \noindent {\em Proof: }
 Choose coordinates $(x^i)$ on $P$ and induced coordinates $(x^i,y^i)$ on
 $TP$. We define a linear operator $T$ on power series (divisible by
$y)$ in $(x,y)$
 by its action on monomials in $y$:
 \[ T y^k \phi(x)  \dgleich \frac{1}{k} y^k \frac{d}{dx} \phi(x) \]

\noindent We will show:
 \begin{enumerate}
\item  \label{pr:anal1}
 The power series of the components $X^i_j$ of
  $X = X^i_j \frac{\partial}{\partial y^i} d x^j$ are majorized
 by
\newline
$f(\sum x^i, \sum y^i)$, where $f(x,y)$ is a power series with no terms
of order zero in $y$ which is a solution of:
\begin{equation} \label{eq:major}
 f(x,y) =  A ( \frac{y^2}{(1-\beta x)(1-\beta y)}  + y T f(x,y)
   + \frac{y}{1-\beta x} f(x,y)  + f(x,y)^2).
\end{equation}
for sufficiently large $A>0$  and $\beta>0$.
\item \label{pr:anal2} $f(x,y)$ is convergent in a neighborhood of $(0,0)$.
\end{enumerate}
Hence, $X$ is analytic.

To show \ref{pr:anal1}, we first observe that Equation $(\ref{eq:major})$ has a
 unique solution containing no term of order zero in $y$, which can be
 computed by iteration.

 We say that $f$ majorizes $X^i_j$ up to  order $k$ in $y$ iff the absolute
value of the coefficient  of any monomial $x^\mu y^\nu$ in $X^i_j$ for
arbitrary multiindices $\mu, \nu$ with
 $|\nu| \leq k$ is not larger than the respective coefficient in
 $f(\sum x^i, \sum y^i)$.

 Both $X^i_j$ and $f(\sum x^i, \sum y^i)$ contain no terms of order zero or one
 in $y$.  As the coefficients of any analytic function
  $\rho(z) = \sum_{\mu \in \natnum^N} C^\mu z^\mu$  in variables
$(z^1,\ldots,z^N)$
 have to satisfy $|C^\mu| < A B^{|\mu|}$ for some constants $A,B$, the function
$\rho$ is
 majorized by $\alpha/(1-\beta (z^1+...+z^N))$ for some $\alpha, \beta>0$. From
 this general statement it follows that there are $\alpha, \beta$ such
 that the components of $\delta a + \delta^{-1} R $ are majorized
 by $\alpha y^2 (1-\beta x)^{-1} (1-\beta y)^{-1}$ , and the
  Christoffel symbols $\Gamma^i_{kl}$ are majorized by
 $\alpha  (1-\beta x)^{-1} $, where $x=\sum x^i, y=\sum y^i$.  Hence,
from equations $(\ref{eq:iter})$ and $(\ref{eq:major})$ it is obvious that for
$A > \alpha$ the power series  $f(\sum x^i, \sum y^i)$ majorizes all
 $X^i_j$ up to order 2 in $y$.

Now assume that $f(\sum x^i, \sum y^i)$ majorizes all $X^i_j$ up to some order
$k \in \natnum$. Then, by construction   of $T$,
the components of $\delta^{-1} d X $ are majorized up to order $(k+1)$ by
$ 2 N y T f(x,y)$, where $N=\dim P$. The components of $ 1/2 \comm{X}{X}$ are
 majorized by  $2 N^2 f(x,y) \frac{d}{dy}f(x,y)$ up to order $k$.
Thanks to the
factor  $1/(p+q)$ in the definition $(\ref{eq:deltainv})$ of $\delta^{-1}$, the
 components of $\delta^{-1}  1/2 \comm{X}{X}$ are majorized up to order $(k+1)$
 by
 \[4  N^3 \int_0^y f(x,\hat{y}) \frac{d}{d\hat{y}}f(x,\hat{y}) d\hat{y}
 = 2 N^3  f(x,y)^2, \]

 Finally, the components of
 $\comm{ \Gamma^i_{kl} y^k \frac{\partial}{\partial y^i} dx^k}{X}$
 are majorized up to order $k$ by $N^2 \alpha /(1-\beta x) (f + y \frac{d}{dy}
f)$ and
 $\delta^{-1}\comm{ \Gamma^i_{kl} y^k \frac{\partial}{\partial y^i} dx^k}{X}$
 is majorized up to order $(k+1)$ by
 \[ 2 N^3  \int_0^y \alpha/(1-\beta x) (f(x,\hat{y}) +
  \hat{y} \frac{d}{d\hat{y}} f(x,\hat{y}) ) d \hat{y}  =
  2 N^3 \alpha /(1-\beta x)  y  f \]
Hence, by equations $(\ref{eq:iter} )$, $(\ref{eq:major})$, $X^i_j$ is
 majorized up to order $(k+1)$ in $y$ by  $f(\sum x^i, \sum y^i)$,
 and \ref{pr:anal1} is shown.

To prove \ref{pr:anal2} we show, again by induction on the degree in $y$,
that $f$ is majorized  by the solution $g(x,y) $  which has no terms of
degree zero  or one in $y$ of the purely algebraic equation
\begin{equation} \label{eq:major2}
 g(x,y) =  \tilde{A} ( \frac{y^2}{(1-\beta x)(1-\beta y)}  +
     2  \frac{y}{1-\beta x} g(x,y)  + g(x,y)^2),
\end{equation}
where $\tilde{A}=\beta A$ if $\beta >1$, and $\tilde{A}=A$ otherwise.
Obviously, $g$ majorizes $f$ up to order $2$. Furthermore, by iteration
of equation $(\ref{eq:major2})$,  $g$
 has the property that the coefficient of $y^k$ consists
 of a sum of terms of the form $a/(1- \beta x)^l$ with
 $a\in \real^+, l \in \natnum ,l\leq k$.
Assume that $f$ is majorized by $g$ up to order $k$ in $y$.
As
\[ y T \frac{y^k}{(1-\beta x)^l} = \frac{l}{k} \frac{\beta y}{1-\beta x}
 \frac{y^k}{(1-\beta x)^l} \]
 $y T f(x,y) $ is majorized up to order $(k+1)$
by $ \beta y g(x,y)/(1-\beta x)$, and hence, by equations
 $(\ref{eq:major})$, $(\ref{eq:major2})$, $f$ is majorized by
$g$ up to order $(k+1)$, so $f$ is majorized by $g$.

 The solutions to the quadratic equation $(\ref{eq:major2})$ are
 \[ g(x,y) = - \frac{1}{2} \left( 2 \frac{y}{1-\beta x} - \frac{1}{\tilde{A}}
\right)
    \pm \sqrt{ \frac{1}{4}  \left( 2 \frac{y}{1-\beta x} - \frac{1}{\tilde{A}}
\right)^2
  - \frac{y^2}{(1-\beta x) (1-\beta y)  }}. \]
 Both solutions are obviously analytic functions in $(x,y)$, as the argument
 of the square root is an analytic function with nonvanishing zeroth degree
 term,
 the solution with the minus sign being the desired solution,  which starts
 with a term quadratic in $y$. Hence $f$ is majorized by a convergent
power series, and \ref{pr:anal2} is shown.
 \hfill Q.E.D. ~~~~\\
\medskip

Since the Ehresmann connection defined by $X$ is flat, the horizontal
 distribution is integrable. Due to the $\delta$-term in $(\ref{eq:Xansatz})$,
 this distribution is transversal to the zero section $Z$, and
  for a  sufficiently small tubular neighborhood
 $U \subset TP$ of $Z$  each leaf in $U$ intersects $Z$
in exactly one point. Hence, we can define an
 ``exponential mapping'' $\EXP_p: T_pP \cap U \rightarrow P$
 for any $p \in P$ which maps $v_p $ to the unique intersection point
 of the leaf through $v_p$ and the zero section in $TP$ (Figure 2).

\begin{figure}
\centering
\mbox{\psfig{file=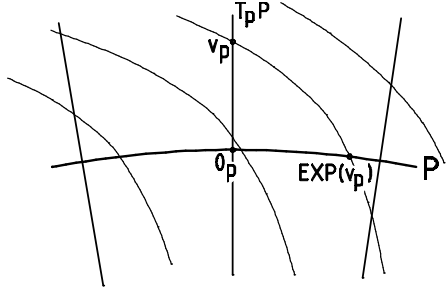,width=6cm,angle=270} }
\caption{The ``exponential mapping'' for a flat linear connection is
found by following parallel sections from the fibres to the zero section.}
\end{figure}

 With this definition, the normalization condition $(\ref{eq:bound})$ with
$a=0$ is
 distinguished by the following theorem:

 \begin{theorem} \label{thm1}
 The mappings $\EXP_p$ and the usual exponential mappings $\exp_p$
 coincide on $U \cap T_pP$ for all $p \in P$ iff the choice
 $\delta^{-1}(X) = 0$ is made in (\ref{eq:bound}).
 \end{theorem}

\noindent {\it Proof:}
 To prove the theorem, we first observe that $X$ determines a natural notion
 of autoparallel curves: Any curve $\gamma$ in $P$ may be lifted to
 a curve in $TP$ by taking the derivative $\dot{\gamma}$. A curve $\gamma$
 is called autoparallel iff $\dot{\gamma}$ is horizontal, i.e., tangent
 to the horizontal distribution defining the connection.

 In local coordinates, this condition is of the form:
 \begin{equation} \label{eq:dgl}
  \ddot{x}^i(t) = - \dot{x}^i(t)  - \Gamma^i_{k l}(x)
      \dot{x}^k \dot{x}^la + X^i_k(x,\dot{x}) \dot{x}^k ~~,
 \end{equation}
 where $ X^i_k(x,y) \frac{\partial}{\partial y^i} dx^k $ is
 the local coordinate expression for $X$.

The connection between autoparallel curves and the `exponential
mapping' $\EXP$ is particular simple in the case of $P=\real^n$
with the standard flat connection $\partial$ (Figure 1). In this case,
the equation for autoparallel curves is just $ \ddot{x}^i(t) = - \dot{x}^i(t)$
with the solution
\[ \dot{x}(t) = \dot{x}(0) e^{-t}, ~~~ x(t) = x(0) + \dot{x}(0) (1 - e^{-t}).\]
The term $- \dot{x}^i(t)$ coming from the $\delta$-term in
$X_{\tot}$ can be interpreted as a friction term slowing the particle
down as it approaches the zero section (Figure 1). In particular,
$EXP(x(0),v(0)) = \lim_{t \rightarrow \infty} x(t) = x(0) + v(0)$, and $EXP$
coincides with the usual exponential mapping of the connection $\partial$.
In the general case, the interpretation of the first term on the right
hand side of $(\ref{eq:dgl})$ as a friction term  is still valid.

 Note that the last term in $(\ref{eq:dgl})$ is precisely $\delta^{-1}(X)$.
Suppose that this is zero. Then  a reparametrization $t \leadsto  -
\log(1-\tau)$
 yields the usual  geodesic equation  corresponding to the connection
  $\partial$ for $\tilde{x}(\tau) \dgleich x(-\log(1-\tau))$.
  In particular, for sufficiently small initial
 velocity $v$, $\tilde{x}(\tau)$ is defined for $0 \leq \tau
 \leq 1$, and $x(t)$ is defined for $0 \leq t < \infty$.

 As $\lim_{t \rightarrow \infty} \tau(t) =\lim_{t \rightarrow \infty}
(1-e^{-t})
 =1$, we get:
 $\lim_{t \rightarrow \infty} x(t)= \lim_{\tau \rightarrow 1} \tilde{x}(\tau)
  = \exp(v)$ with $v=  \dot{x}(0)$.
 On the other hand,   by construction the curve $(x,\dot{x})(t)$ in $TP$
  is contained in the leaf of the horizontal foliation through $v$. As
 $\lim_{t \rightarrow \infty} \dot{x}(t) = \lim_{t \rightarrow \infty}
 e^{-t} \frac{d}{d\tau} \tilde{x}(\tau) = 0$, $\lim_{t \rightarrow \infty}
x(t)$
 is just the intersection point of the leaf through $v$ and the zero
 section, i.e. $\EXP(v)$. Hence, $\EXP(v)=\exp(v)$ and we
have proved that the condition $\delta^{-1}(X)=0 $ is sufficient
for the maps $\EXP$ and $\exp$ to coincide.

To prove the converse we
 assume $\EXP_p = \exp_p $ for all $p \in P$, and
 $\delta^{-1} X$ arbitrary. We first show the following lemma:

 \begin{lemma} \label{lem1}
 Let $\gamma$ be a geodesic corresponding to the connection $\partial$,
  $\tilde{\gamma} : \real \rightarrow P  $ the reparametrized geodesic:
   $ t \mapsto \gamma(1-e^{-t})$, i.e., a solution to the differential
 equation
 \begin{equation} \label{eq:dgl_noX}
  \ddot{x}^i(t) = - \dot{x}^i(t) - \Gamma^i_{k l}(x)
      \dot{x}^k \dot{x}^l .
 \end{equation}

  Then, for all $t$:
 \[ \exp_{\tilde{\gamma}(t)} \left( \dot{ \tilde{\gamma}}(t) \right)
   = \exp_{\tilde{\gamma}(0)} \left( \dot{ \tilde{\gamma}}(0) \right)
 = \lim_{t \rightarrow \infty} {\tilde{\gamma}(t)} \]
 \end{lemma}

 \noindent {\bf Remark:}  The lemma says that a solution of
$(\ref{eq:dgl_noX})$
 is a reparametrized geodesic that is slowed down by the `friction term'
 $- \dot{x}$ just to that extent that $\exp(\dot{\tilde{\gamma}}(t))$ is
 constant.
 \bigskip

 \noindent {\em Proof of the lemma:}

 Denote by $\gamma_2$ the geodesic starting at $\tilde{\gamma}(t_0)$
 with initial velocity $\dot{\tilde{\gamma}}(t_0)$ for some fixed time $t_0$.
 As follows from the first part of the proof of theorem \ref{thm1}
 above, $\tilde{\gamma}_2(t)= \gamma_2(1-e^{-t})$ defines again a solution
 of $(\ref{eq:dgl_noX})$. Obviously, this solution is obtained from
 $\tilde{\gamma}$ by a constant shift in time, as
  $ \dot{\tilde{\gamma}}_2(0)= \dot{\tilde{\gamma}}(t_0)$.
 Hence,
 $\lim_{t \rightarrow \infty} \tilde{\gamma}(t) = \lim_{t \rightarrow \infty}
 \tilde{\gamma}_2(t)$. On the other hand, we have already proved that
 $\lim_{t \rightarrow \infty} \tilde{\gamma}(t)
 = \exp_{\tilde{\gamma}(0)}(\dot{\tilde{\gamma}}(0))$ (and hence
 $\lim_{t \rightarrow \infty} \tilde{\gamma}_2(t)
 = \exp_{\tilde{\gamma}_2(0)}(\dot{\tilde{\gamma}}_2(0))$), so the
 lemma follows. \hfill Q.E.D. ~~~~\\ \medskip

Using this lemma we will finish the proof of Theorem \ref{thm1}, i.e.,
we will show that the condition $\delta^{-1} X = 0$ is a necessary condition
for the maps $\EXP$ and $\exp$ to coincide.

Assume that the curve $t \mapsto \frac{d}{dt} \gamma(1-e^{-t}) $ in
 $TP$ is not completely contained in the leaf through $v_q$, where
 $\gamma: t \mapsto \gamma(t)$
is a geodesic of $\partial$
with $\gamma(0) = v_q $ for an arbitrary $v_q \in TP$. By continuity
 there is a open subset $O \subset \real$ such that $w(t) := \frac{d}{d t}
 \gamma(1-e^{-t}) $ is not contained in the leaf through $v_q$ for any
 $t \in O$. By assumption and the lemma, $\EXP(w(t)) = \exp(w(t)) = \exp(v_q)
 = \EXP(v_q)$, so we get an infinity of different leaves intersecting the
 zero section in the same point, in contradiction to the transversality
 of the horizontal distribution.

 Hence, the curve $t \mapsto \frac{d}{dt} \gamma(1-e^{-t}) $ is completely
 contained in the leaf through $v_q$, i.e., it is a horizontal curve,
 and therefore must fulfill both equations $(\ref{eq:dgl})$ and
 $(\ref{eq:dgl_noX})$. Obviously, this is only possible if the
 difference between the two equations vanishes, i.e.,
 \[ X^i_k(x,\dot{x}) \dot{x}^k \equiv 0 \Leftrightarrow
 \delta^{-1} X = 0 ,\]
and Theorem \ref{thm1} is proved.
 \hfill Q.E.D. ~~~\\ \medskip

 \noindent {\bf Remark: }  Even when $\delta^{-1} X \neq 0$, the equality
 $\EXP(v) = \lim_{t \rightarrow \infty}  x(t)$, where $x(t)$ is the
  solution of $(\ref{eq:dgl})$
 with $\dot{x}(0)=v$, still holds. To prove this statement we  show that
 the `friction term' in $(\ref{eq:dgl})$ insures that
 $\lim_{t \rightarrow \infty}  \dot{x}(t)=0$ for sufficiently small initial
 velocity $v$. However, this is easy to see: Multiplying $(\ref{eq:dgl})$ by
 $\dot{x}^i$ and summing over $i$, we get:
 \[ \frac{1}{2} \frac{d}{d t}\sum_i \dot{x}^i \dot{x}^i  =
  -\sum_i \dot{x}^i \dot{x}^i -
  \sum_i \dot{x}^i \left( \Gamma^i_{k l}(x)
      \dot{x}^k \dot{x}^l + X^i_k(x,\dot{x}) \dot{x}^k \right)
 \]

 As $X$ has degree at least two in $y$, for sufficiently small velocity
 \[ | \sum_i \dot{x}^i \left( \Gamma^i_{k l}(x)
      \dot{x}^k \dot{x}^l + X^i_k(x,\dot{x}) \dot{x}^k \right)| <
  \epsilon \sum_i \dot{x}^i \dot{x}^i \]
 for some $\epsilon$ with  $0<\epsilon <1$.

 Hence,
 \[ (-1- \epsilon) \sum_i \dot{x}^i \dot{x}^i  < \frac{1}{2} \frac{d}{dt}
   \sum_i \dot{x}^i \dot{x}^i
 (-1 +\epsilon) \sum_i \dot{x}^i \dot{x}^i, \]

 \[ e^{-2(1+ \epsilon)t} \sum_i \dot{x}^i(0) \dot{x}^i(0)  <
  OA \sum_i \dot{x}^i(t) \dot{x}^i(t)
 <  e^{-2(1-\epsilon)t} \sum_i \dot{x}^i(0) \dot{x}^i(0), \]
 and
 \[ \lim_{t\rightarrow \infty} \dot{x}(t) =0 \]
but the curve $(x(t),\dot{x}(t))$ does not intersect the zero section
at a finite time.

\section{ The formal exponential map}
In the last section we have shown how to construct the mapping $\EXP$
in the analytic case, where the formal flat connection was convergent
and defined a flat Ehresmann connection in the usual sense on some
neighborhood of the zero section. In the non-analytic case we
cannot expect the series for $X$ to be convergent. However, we will
show that we can define a formal exponential map, i.e., an $\infty$-jet
of mappings from the fibre $T_pP$ to $P$, and we will show that for this map
Theorem \ref{thm1} still holds, if we replace $\exp_p$ by
its $\infty$-jet as well.

In order to define the formal exponential map we first show
the following lemma for real-analytic manifolds:

\begin{lemma} \label{lem:jets}
 Let $P$ be a real-analytic manifold, $\partial$ an
analytic connection and $a$ an analytic function, $X$ the solution
of $(\ref{eq:curvzero})$ with the normalization condition $(\ref{eq:bound})$.
 Then the $k$-jet  of $\EXP_p$ at $0_p \in T_pP$ for arbitrary
 $k \in \natnum, p \in P$ is completely
determined by the $k$-jet at $p$ of the connection form corresponding to
$\partial$ and the $(k+1)$-jet at $0_p$ of $a$.
\end{lemma}

\noindent {\em Proof: }
It is sufficient to consider some small neighborhoods $V\subset P$ of
 $p \in P $ and $TV \subset TP$ of $0_p \in T_pP$.
We choose coordinates $(x^i)$ on $V$ and  show  first that the $k$-jet of
$\EXP_p$ is determined by  the $k$-jet of $X$.

By the definition of the map $\EXP_p$, we may compute the components $\EXP_p^i$
 of $\EXP_p$ in the chosen coordinate system by taking the covariant constant
continuation $f^i$ to $TV \subset TP$ of the coordinate functions $x^i$ and
 restricting them to the fibre $T_pP$:
\[ \EXP_p^i(y) = f^i(y) \restr{T_pP} \]
with
\[ D f^i = 0 , ~~~~~ f^i\restr{Z} = x^i ,\]
where $Z$ denotes the zero section in $TP$.

As the constructed connection is flat, we know that such functions $f^i$
exist.  Rather than constructing them by the iterative procedure of
\cite{fe:simple}, we will use a more geometric approach.
The condition $D f^i = 0$ reads locally:
\begin{equation} \label{eq:Dexp}
 \frac{\partial}{\partial x^k} f^i +
 M^m_k  \frac{\partial}{\partial y^m} f^i  = 0,
\end{equation}
where $M$ is the matrix with elements
\begin{equation} \label{eq:defM}
 M^m_k = \delta^m_k + \Gamma^m_{jk} y^j -X^m_k,
\end{equation}
where $\delta^m_k $ equals $1$ if $m=k$, and $0$ otherwise.
Due to the $\delta^m_k$ coming from the solder form the matrix valued
function  $M$ is invertible,
and the inverse $M^{-1}$ is analytic at $0_p$ again. Furthermore, its
$k$-jet at $0_p$ only depends on the $k$-jet of $M$ at $0_p$.

Hence, we can solve (\ref{eq:Dexp}) for $ \frac{\partial}{\partial y^k} f^i$
yielding
\begin{equation} \label{eq:Dexpinv}
 \frac{\partial}{\partial y^k} f^i +
 (M^{-1})^m_k  \frac{\partial}{\partial x^m} f^i  = 0,
\end{equation}
where the zero-jet of $M^{-1}$ at $0_p$ is just the unit matrix.
$f^i$ is given for $y=0$, and we know that this set of partial
differential equations has a unique solution, since the connection is
flat. To compute it,
we can integrate the differential equations successively:
The equation for $ \frac{\partial}{\partial y^1} f^i$ determines
$f^i$ on the $y^1$-axis, the equations for $\frac{\partial}{\partial y^k} f^i$
with $1\leq k\leq m$ determine $f^i$ on the subspace spanned by the
first $m$ basis vectors. In each step the solution is an analytic
function on the respective subspace, depending analytically on
the initial conditions, by the Cauchy-Kovalevskaya theorem.
 Hence, $f^i$ is analytic at $0_p$.
Furthermore, by comparing coefficients, we see that the
$k$-jet of $f^i$ only depends on the $k$-jet of $M^{-1}$, and hence
of $M$.

Since the operator $\delta^{-1}$ contains a factor $y^i$, it follows
from the iteration formula $(\ref{eq:iter})$ for $X$ that the
$k$-jet of $X$ is determined by  the $k$-jets of the Christoffel-symbols
$\Gamma^i_{jk}$ of $\partial$ and the $(k+1)$-jet of $a$.
By the definition $(\ref{eq:defM})$ this is true for $M$ as well, and
the lemma is proved. \hfill Q.E.D. ~~~~\\ \medskip

Using this lemma, we can define a formal exponential map in the non-analytic
case: We use the topology on the $\infty$-jets generated by the basis of
open sets consisting of sets of $\infty$-jets whose $k$-jets agree for some
$k \in \natnum $,
i.e., a sequence $(j^i)$ of $\infty$-jets converges to an $\infty$-jet $j$,
if for any $k \in \natnum$ there is an $N(k) \in \natnum$ auch that the
$k$-jets of $j^r$ and $j$ agree for $r > N(k)$.

In this topology the set of $\infty$-jets of analytical  functions
forms a dense subset of the set of all $\infty$-jets. (The polynomials
are already dense.)  Hence,
 we can define  the $\infty$-jet   $\jEXP$ of $\EXP$ at $p$ in the
 non-analytic case by  the continous continuation to all
$\infty$-jets  of the map that assigns to the pair $(\partial,j^\infty_p(a))$,
 consisting of an analytic connection and  the $\infty$-jet of an analytic
function $a$ defining the normalization condition for $X$,
 the $\infty$-jet  $\jEXP$.
This continuation exists and is unique by Lemma \ref{lem:jets}.

\begin{theorem}
Theorem \ref{thm1}  still holds in the non-analytic case,
if we replace $\EXP_p$, and $\exp_p$ by their $\infty$-jets.
\end{theorem}

\noindent {\em Proof:}
  If we impose the normalization condition $(\ref{eq:bound})$
with $a=0$ the direct statement of the theorem is an immediate
consequence of Theorem
\ref{thm1}, Lemma \ref{lem:jets}, and the fact that the $\infty$-jets
of analytic
functions are dense in the chosen topology.

To show the converse, we use the fact that $j^\infty_p(\exp_p)$
coincides with $\jEXP$ for the choice $a=0$ in  $(\ref{eq:bound})$
by the first part of the theorem.
 We denote for the moment the corresponding $X$ by $X^0$.
Now, assume that the $k$-jet of $a$ in $(\ref{eq:bound})$ is different
from zero for some $k >2$. Then, by $(\ref{eq:iter})$, the
$(k-1)$-jet of $X$ and $X^0$ are different as well. This implies,
that the $(k-1)$-jets of $M$ and $M^{-1}$ in $(\ref{eq:Dexp}),
(\ref{eq:Dexpinv})$ are different
from those for $a=0$, so the $(k-1)$-jets of the corresponding
jets $\jEXP$ are different.   \hfill Q.E.D. ~~~~\\ \medskip

\section{ Flat  symplectic connections}
\label{sec:sympl}
So far, we have studied the problem of constructing a flat connection
for an arbitrary manifold $P$, without any additional structure. However,
if we are interested in getting closer to the quantum case,
 then $P$ should be a classical
phase space with its symplectic
structure, and ${\cal F}(P)$ a Poisson algebra.

In this case, the Weyl-algebra bundle of \cite{fe:simple} is replaced by
 ${\cal A}= {\cal F}(TP)$
 (or  ${\cal A} = \Gamma( \bigcup_{p \in P} {\cal J}^\infty_0(T_p P,\real))$)
with the fibrewise Poisson-structure $\{,\}_{\fib}$. The flat connection
 must be compatible with this additional structure,
 i.e., the following equation should hold for all $f,g \in {\cal A}$:
 \begin{equation} \label{eq:Dderiv}
    D\{f,g\}_{\fib} = \{ D f, g \}_{\fib} + \{ f , D g\}_{\fib}.
 \end{equation}

 Hence, $\partial $ has to be a symplectic connection and $X(v)$ has to
 be a symplectic vector field on $T_{x}P$ for any $v \in T_{x}P$.
  As the fibres are
 vector spaces, $X$ must be globally hamiltonian, i.e., there must
 be an ${\cal A}$-valued one-form $r$ such that the components of
 the one form  $r$  are minus the hamiltonian functions of the components
of the vector-valued one-form $X$.
 (The minus sign is chosen
 in order to guarantee that the Lie bracket of $X$ with itself corresponds
  to the Poisson bracket of $r$ with itself.)

 In this case, the normalization condition $\delta^{-1} X = 0$ is in
 general not admissible, as it will not give a symplectic vector
 field. This statement is an immediate consequence of the fact that the
 usual exponential mapping of a symplectic connection is
 generally not a local symplectomorphism, and the following theorem:

 \begin{theorem}
 The mapping $\EXP(p)$
  is a local symplectomorphism iff $X(v)$ is
 a hamiltonian vector field for any $v \in TP.$
 \end{theorem}

 \noindent {\em Proof: }
 Let $X(v)$ be a hamiltonian vector field for any $v \in TP$, with
 $X = X_r$ for some ${\cal A}$-valued one-form $r$.
 Then
  $D$ is of the form:
 \[ D a = \partial a + \{ \omega_{ij} y^i dx^j,a\}_{\fib} + \{r,a\}_{\fib} ,\]
 and  has the property $(\ref{eq:Dderiv})$.

 We have to show:
 \begin{equation} \label{eq:sympl}
  \EXP_p^* \{f,g\} = \{\EXP_p^* f, \EXP_p^* g\}_p
 \end{equation}
 for all $f,g \in {\cal F}(P)$,
 where $\{,\}_p$ is the Poisson bracket on the fibre $T_p P$ induced by
 the constant symplectic form $\omega(p)$, i.e., the restriction of
 $\{,\}_{\fib} $ to the fibre $T_pP$.

 For any $ a \in {\cal F}(P)$ let $\sigma(a)$ denote the unique covariant
constant
 continuation of $a$, i.e., $\sigma(a) \in {\cal A}$, $D \sigma(a) =0,
 \sigma(a) \restr{P \subset TP} =a$, where we have identified $P$ with
 the zero section in $TP$.

 By construction of $\EXP$ we have:
 \begin{equation} \label{eq:exp}
  \EXP_p ^* f = \sigma(f) \restr{T_p P} .
 \end{equation}
 Obviously, due to the $\delta$-term in $X_{\tot}$,
 $\sigma(a)$ is locally of the form:
 \[\sigma(a)(x,y) = a(x) + \frac{\partial}{\partial x^i}a(x)  y^i + O(y^2) \]
 and hence:
 \[ \{ \sigma(f), \sigma(g) \}_{\fib} \restr{P} = \{ f, g \}, ~~~~~
  D\{ \sigma(f), \sigma(g) \}_{\fib} = 0~~~~~~~(\mbox{by
(\ref{eq:Dderiv}))} \]
 \[\Rightarrow \{ \sigma(f), \sigma(g) \}_{\fib} = \sigma(\{f,g\}) \]
 for any $f,g \in {\cal F}(P)$ by uniqueness of the covariant constant
 continuation of a function in ${\cal F}(P)$.
 Using $(\ref{eq:exp})$, equation $(\ref{eq:sympl})$ follows.
 \hfill Q.E.D. ~~~~~\\ \medskip

For a given linear symplectic connection
 the set of  its nonlinear ``flattenings'' satisfying $(\ref{eq:Dderiv})$
is in one-to-one correspondence
 with the functions $ f \in {\cal A}$ at least of degree  4 in the fibres
  by the condition:
  $\delta^{-1} r = f$, where $r$ is the unique one form
 in ${\cal A} \tensor \Lambda(P)$ which is
 a hamiltonian for $X$ and vanishes on $P \subset TP$.
 Indeed, as any $X$ is at least of second degree in the vertical coordinates,
 $r$ is at least of third, and $\delta^{-1} r$ at least of fourth degree.

 On the other hand, given $f$, we may write the analog of equation
 $(\ref{eq:iter})$ for $r$ instead of
 $X$, which is essentially the same (up to signs, since X is minus the
 hamiltonian vector field of $r$, and the replacement of the commutator
 of  vector fields by the Poisson bracket of functions.)
 Hence, given $f$, we can
 uniquely construct $r$ as the  solution of the equation:
 \begin{equation} \label{eq:rcurv}
  0 = \hat{R} - \delta r + \partial r +  1/2\{r,r\} .
 \end{equation}
 with the condition
 \begin{equation} \label{eq:rcond}
 \delta^{-1} r = f,
 \end{equation}
 where the ${\cal A}$-valued two-form  $\hat{R}$ is the hamiltonian for the
 curvature $R$ considered
 as a two form on $P$ with values in the vertical vector fields on
 $TP$, locally given by
 $\hat{R} = (1/4) \omega_{ij} R^j_{klm} y^i y^k d x^k d x^m $.
 This solution may again be iteratively computed:
 \begin{equation} \label{eq:riter}
  r = \delta f + \delta^{-1}(\hat{R} + \partial r + 1/2 \{r,r\}).
 \end{equation}

 We observe
 that the connection on $U \subset TP$  can be easily reconstructed
 from the mapping $\EXP$: The leaf through any $0_p \in TP$ is given
 by $\bigcup_{q \in P} \EXP_q^{-1}(\{0_p\})$, and the connection is
 defined by the tangent distribution to those leaves. Since the mappings
 $\EXP_p$ for different symplectic connections $\partial$ and different
 `normalization conditions' $(\ref{eq:rcond})$ are all local
symplectomorphisms, this observation implies the following theorem:

 \begin{theorem}
 The different Ehresmann connections constructed from different
 symplectic connections and different choices of $f$ are related
 by fibrewise symplectomorphism, i.e., by a section in the bundle
 $\bigcup_{p \in P} \mbox{\it Sym}^{\,0_p}(T_pP)$ where
 $\mbox{\it Sym}^{\,0_p}(T_pP)$ denotes the group of
(infinite jets of) symplectomorphisms
 of $T_pP$ with fixed point $0_p \in T_pP$.
 \end{theorem}

This statement is the symplectic analog of Theorem 4.3 in \cite{fe:simple}
for the quantum case, which states that two different `abelian connections'
$D, \tilde{D}$ are related by some inner automorphism $U$ such that
$\tilde{D} = D - [ D U \circ U^{-1}, \cdot ]$. (The theorem is
formulated only for the normalization condition $\delta^{-1} r = 0$.
However, this condition is not used in the proof, so it is valid for
the more general normalization conditions studied by us as well.)

 For the condition $\delta^{-1} r = 0$, $\EXP$ is in a suitable sense
a `minimal deformation' of the usual exponential mapping such that the
resulting mapping is symplectic: namely, as the symplectic connection
$\partial$ defines a splitting of $TTP$ in horizontal and vertical
subspaces, we can define the accelaration $a(v_p)$ of autoparallel
curves due to $X$ and the friction term $- \delta$ at a point $v_p \in
TP$ as the difference of the tangent vectors to the lifted
autoparallel curve and the lifted geodesic corresponding to $\partial$
through that point, which is always a vertical vector. We identify
$T(T_p P)$ and $T_pP $, and denote by $\uparrow$ the vertical lift:
$\uparrow: T_p P \rightarrow T(T_p P), ~~ v\mapsto v^{\uparrow}$.  In
order to avoid notational confusion, we denote by $\tilde{\omega}_p $
the symplectic form on the vector space $T_pP$ induced by the
symplectic form $\omega$ on $P$.

 \begin{theorem}
 The condition $\delta^{-1} r = 0$ for a solution of $(\ref{eq:rcurv})$
 is equivalent to the requirement that the acceleration $a(v_p)$ is contained
 in the $\tilde{\omega}_p$-complement of $v_p^{\uparrow}$ for all
$p \in P, v_p \in T_pP$,
 i.e, $\tilde{\omega}_p(v_p^\uparrow, a(v_p)) = 0 ~\forall v_p \in TP$.
 \end{theorem}

 \noindent {\it Proof: }  Any solution $r$ of $(\ref{eq:rcurv})$ is of the
 form $(\ref{eq:riter})$ for some $f$. $a(v_p)$ may be written as
\[ a(v_p) = - v_p^{\uparrow} + a^X(v_p), \]
where the first term comes from the friction term $- \delta$ and $a^X(v_p)$
is the acceleration due to $X$.   For arbitrary $f$, we conclude from
 $(\ref{eq:dgl})$ that $a^X(v_p)$ is just $(\delta^{-1} X)(v_p)$,
 which is nonzero even if  $\delta^{-1} r =0$, as  $\delta^{-1}$ explicitly
 contains the vertical coordinates $y$.

 Using the definition of $\delta^{-1}$ we get:
 \begin{equation} \label{eq:delXr}
  \delta^{-1} X
 =  \frac{I+1}{I} X_{\delta^{-1} r}  + (\omega^\#( I^{-1} r))^{\uparrow},
 \end{equation}
 where $X_f$ denotes the hamiltonian vector field on the fibres of $TP$
  corresponding to $f$ (i.e., $\tilde{\omega}^\#(d_y f)$),
 and $\frac{I+1}{I}, I^{-1}$ are linear operators acting on monomials
 in $y$ of total degree $i$ by multiplication with
 $\frac{i+1}{i}$ and $ i^{-1}$, respectively.

 By $(\ref{eq:riter})$ $r$ is of the form $r = \delta f + \delta^{-1} \alpha$
 for a  two form $\alpha \in {\cal A} \tensor \Lambda(P)$.

 Setting $V_p = \sum y^i \frac{\partial}{\partial x^i}$ and using
  $(\ref{eq:delXr})$ we get:
 \begin{eqnarray*}
  \tilde{\omega}_p(\delta^{-1} X , V_p^{\uparrow})  &=&
  \tilde{\omega}(\frac{I+1}{I} X_{f}, V_p^{\uparrow})
   + \tilde{\omega}(\omega(p)^\#(  I^{-1} (\delta f + \delta^{-1}
\alpha))^{\uparrow} , V_p^\uparrow)\\
 &=& (\delta f)(V_p) +  (\frac{1}{I} \delta^{-1} \alpha)(V_p) \\
 &=&  (\delta f)(V_p) ,
 \end{eqnarray*}
 where the last equality holds because for each monomial in $y$,  the
expression
 $ (\delta^{-1} \alpha)(V_p)$ is --- up to a real factor --- just
 $\alpha(V_p,V_p) = 0$.
 Hence,  $\tilde{\omega}_p(v_p^\uparrow, a(v_p)) = 0~~ \forall v_p \in TP$ is
 fulfilled for $f \equiv 0$.

 On the other hand, since
  $ (\delta f)(v_{p}) = \frac{d}{d \lambda} f(x,\lambda y)\restr{\lambda=1} $,
  it follows from the fact that $$\tilde{\omega}_p(v_p^{\uparrow},
a(v_p)) = 0~~ \forall v_p \in TP$$
 that $f(x,y)$ has to be independent of $y$ and hence zero, as the
 constant term in $f$ must vanish.
 \hfill Q.E.D. ~~~~ \\

\end{document}